\documentclass[10pt,draftcls, onecolumn]{IEEEtran}
\usepackage{amssymb, color, bm, cite, psfrag, rotating, url}
\usepackage{graphicx}
\usepackage{epstopdf}
\AppendGraphicsExtensions{.png}

\begin{document}
\title{Spread spectrum magnetic resonance imaging}
\author{Gilles~Puy, Jose P. Marques, Rolf~Gruetter, Jean-Philippe Thiran, Dimitri~Van De Ville, Pierre~Vandergheynst and Yves~Wiaux
\thanks{This work is supported in part by the Center for Biomedical Imaging (CIBM) of the Geneva and Lausanne Universities, EPFL, and the Leenaards and Louis-Jeantet foundations, in part by the Swiss National Science Foundation (SNSF) under grant PP00P2-123438, and also by the EPFL-Merck Serono Alliance award.}
\thanks{G.~Puy, J.-Ph.~Thiran, P.~Vandergheynst and Y.~Wiaux are with the Institute of Electrical Engineering,  Ecole Polytechnique F{\'e}d{\'e}rale de Lausanne (EPFL), CH-1015 Lausanne, Switzerland. G.~Puy, J. P. Marques, and R.~Gruetter are with the Institute of the Physics of Biological Systems,  Ecole Polytechnique F{\'e}d{\'e}rale de Lausanne (EPFL), CH-1015 Lausanne, Switzerland. J. P. Marques and R.~Gruetter are with the Department of Radiology, University of Lausanne (UNIL), CH-1015 Lausanne, Switzerland. Y.~Wiaux and D.~Van De Ville are with the Institute of Bioengineering,  Ecole Polytechnique F{\'e}d{\'e}rale de Lausanne (EPFL), CH-1015 Lausanne, Switzerland. Y.~Wiaux, R.~Gruetter, and D.~Van De Ville are with the Department of Radiology and Medical Informatics, University of Geneva (UniGE), CH-1211 Geneva, Switzerland.}
\thanks{E-mail: gilles.puy@epfl.ch; yves.wiaux@epfl.ch - Address: EPFL STI IEL LTS2 - ELE 227 - Station 11 - CH-1015 Lausanne.}} 

\maketitle

%
%
% ABSTRACT
%
%
\begin{abstract}
We propose a novel compressed sensing technique to accelerate the magnetic resonance imaging (MRI) acquisition process. The method, coined spread spectrum MRI or simply s$_2$MRI, consists of pre-modulating the signal of interest by a linear chirp before random $\bm{k}$-space under-sampling, and then reconstructing the signal with non-linear algorithms that promote sparsity. The effectiveness of the procedure is theoretically underpinned by the optimization of the coherence between the sparsity and sensing bases. The proposed technique is thoroughly studied by means of numerical simulations, as well as phantom and \emph{in vivo} experiments on a 7T scanner. Our results suggest that s$_2$MRI performs better than state-of-the-art variable density $\bm{k}$-space under-sampling approaches.
\end{abstract}

\begin{IEEEkeywords}
compressed sensing, spread spectrum, magnetic resonance imaging.
\end{IEEEkeywords}

%
%
% INTRODUCTION
%
%
\section{Introduction}

In magnetic resonance imaging (MRI), the signal of interest $\rho$ represents the magnetization induced by resonance in the imaged tissues. In the standard setting, data acquired in MRI provide complete Fourier, or $\bm{k}$-space, measurements of this signal $\rho$. Accelerating the acquisition process, or equivalently increasing the achievable resolution for a fixed acquisition time, is of major interest for MRI applications. Recent approaches based on compressed sensing seek to reconstruct the signal from incomplete $\bm{k}$-space information, hence defining an ill-posed inverse problem. The ill-posed inverse problem is regularized by the introduction of sparsity priors, acknowledging the fact that many MRI signals are sparse in well-chosen bases; i.e., that their expansion contains only a small number of non-zero coefficients. Such compressed sensing techniques have been developed for static and dynamic imaging \cite{lustig07, jung07, gamper08, jung09, usman11}, parallel MRI \cite{liang09a, lustig10, otazo10}, MR spectroscopic imaging \cite{hu08, hu10, larson10}, and many other applications. Several algorithms have also been proposed to reconstruct MRI signals from under-sampled $\bm{k}$-space measurements (see, e.g., \cite{kim07, trzasko09, kern11}).

In the framework of compressed sensing, signals are usually measured through random matrices to ensure that any sparse signal can be recovered with overwhelming probability. The common approach in MRI consists in also exploiting the fact that the energy of MRI signals is usually concentrated at low spatial frequencies. Therefore a variable density $\bm{k}$-space random sampling where the under-sampling ratio increases at high frequencies is usually used \cite{lustig07}. This method, heuristic in nature, was shown to be very effective in enhancing the signal reconstruction quality when random distributions optimizing the associated point spread function are used. Other approaches that optimize the acquisition procedure have also been proposed: $\bm{k}$-space sampling optimization by Bayesian inference \cite{seeger10}; random $\bm{k}$-space convolution with Toeplitz matrices \cite{sebert08, liang09b, wang09}; encoding by projection onto random waveforms with Gaussian distributions \cite{haldar10}.

In the present work, we propose the use of a spread spectrum technique to accelerate single coil MRI acquisition in the framework of compressed sensing. We study this method, coined spread spectrum MRI or simply s$_2$MRI, theoretically, numerically via simulations, and empirically via real acquisitions. The essence of our strategy consists of pre-modulating the image by a linear chirp, which results from the application of quadratic phase profiles, and then performing random $\bm{k}$-space under-sampling. Images are then reconstructed with non-linear algorithms promoting signal sparsity. The enhancement of the signal reconstruction quality is linked to a decrease of coherence of the measurement system \cite{puy11a, wiaux09b}. In MRI, this type of modulation is known as phase scrambling. It can be obtained by using dedicated coils or by modifying radio frequency (RF) pulses. It has been used for various purposes, such as improving dynamic range \cite{maudsley88, wedeen88}, or reducing aliasing artifacts \cite{pipe95, ito08}, but never in a compressed sensing perspective.

Let us acknowledge that this spread spectrum technique was initially introduced by some of the authors for compressive sampling of pulse trains in \cite{naini09}. Its transfer to a setting encompassing analog signals and modulations was studied in the context of radio interferometry \cite{wiaux09b, wiaux09d, mcewen10}. The effectiveness of the method for MRI was briefly discussed in \cite{puy09, wiaux09c, puy11b}.

The paper is organized as follows. In Section \ref{sec:spread spectrum principle}, we explain the principle of the spread spectrum technique in a simplified analog setting. In Section \ref{sec:MR measurements}, we formulate the inverse problem for image reconstruction from under-sampled $\bm k$-space measurements in the presence of chirp modulation and compare the s$_2$MRI technique to the variable density sampling method on the basis of numerical simulations of MR acquisitions of a brain image. In Section \ref{sec:Real acquisitions}, we describe our implementation of the chirp modulation (on a $7$T scanner) and show the effectiveness of s$_2$MRI using numerical simulations in this precise setting and real acquisitions of phantom and \emph{in vivo} data. Finally, we conclude and discuss potential evolutions of the
technique in Section \ref{sec:Conclusion}.

%
%
% SPREAD SPECTRUM PRINCIPLE
%
%
\section{Spread spectrum principle}
\label{sec:spread spectrum principle}

%
%
% Compressed sensing
%
%
\subsection{Compressed sensing}
\label{sub:compressed sensing}

The essence of the recent theory of compressed sensing is the merging of data acquisition and compression \cite{candes06a, candes06b, candes06c, donoho06, candes07, baraniuk07a, rauhut10, donoho09}. Beyond MRI, it is well-known that a large variety of natural signals are sparse or compressible in multi-scale bases, such as wavelet bases. In the compressed sensing theory, signals are usually expressed as $N$-dimensional vectors: $\bm{\rho}\in\mathbb{C}^{N}$. By definition, a signal is sparse in some orthonormal basis $\mathsf{\Psi}\in\mathbb{C}^{N\times N}$, called a sparsity basis, if its expansion contains only a small number $K\ll N$ of non-zero coefficients. More generally it is compressible if its expansion contains only a small number of significant coefficients. The decomposition of $\bm{\rho}$ in $\mathsf{\Psi}$ is denoted $\bm{\alpha} \in \mathbb{C}^{N}$, and it satisfies
\begin{equation}
\label{cs1}
\bm{\rho} = \mathsf{\Psi}\bm{\alpha}.
\end{equation}

The theory of compressed sensing demonstrates that a small number $M\ll N$ of linear and non-adaptative measurements $\bm{\nu}\in\mathbb{C}^{M}$ suffices for an accurate and stable reconstruction of the signal $\bm{\rho}$. These measurements may, for example, be obtained by projection onto $M$ randomly selected basis vectors of an orthonormal basis $\mathsf{\Phi}\in\mathbb{C}^{N\times N}$, called a sensing basis. This selection process can be modeled by a multiplication with a rectangular binary matrix $\mathsf{M}\in\mathbb{R}^{M\times N}$ that contains only one non-zero value on each line, at the index of the basis vector to be selected. To model a non-perfect sensing process, these measurements are assumed to be contaminated by independent and identically distributed noise $\bm{n} \in \mathbb{C}^{M}$. The measurement model thus satisfies
\begin{equation}
\label{cs2}
\bm{\nu}=\mathsf{\Theta}\bm{\alpha}+\bm{n}\textnormal{, with }\mathsf{\Theta}=\mathsf{M\Phi^{\star}\Psi}\in\mathbb{R}^{M\times N},
\end{equation}
where the matrix $\mathsf{\Theta}$ identifies the measurement matrix as seen from the sparsity basis ($\cdot^\star$ denotes the conjugate transpose operation).

To reconstruct the signal $\bm{\rho}$ from the measurements $\bm{\nu}$, the compressed sensing framework proposes, among other approaches, to solve the Basis Pursuit (${\rm BP}$) minimization problem \cite{candes06a, candes06b, candes06c, donoho06, candes07, baraniuk07a, donoho09, rauhut10}. This problem regularizes the originally ill-posed inverse problem related to (\ref{cs2}) with an explicit sparsity or compressibility prior on the signal. In the presence of noise, the ${\rm BP}$ problem is the minimization of the $\ell_{1}$ norm\footnote{$\vert\vert \bm{\alpha} \vert\vert_{1}=\sum_{i=1}^{N}\vert \alpha_{i}\vert$ where $\alpha_{i}$ is the $i$th entry of the vector $\bm{\alpha}$.} of $\bm{\alpha}$ under a constraint on the $\ell_{2}$ norm\footnote{$\vert\vert \bm{n} \vert\vert_{2}=(\sum_{i=1}^{M}\vert n_{i}\vert^{2})^{1/2}$ where $n_{i}$ is the $i$th entry of the vector $\bm{n}$.} of the residual noise $\bar{\bm{n}}= \bm{\nu}-\mathsf{\Theta}\bar{\bm{\alpha}}$:
\begin{equation}
\label{cs3}
\bm{\alpha}^* = \arg \min_{\bar{\bm{\alpha}}\in\mathbb{C}^{N}}\vert\vert\bar{\bm{\alpha}}\vert\vert_{1}\textnormal{ subject to }\vert\vert \bm{\nu}-\mathsf{\Theta}\bar{\bm{\alpha}}\vert\vert_{2}\leq\epsilon.
\end{equation}
The corresponding reconstructed signal is $\bm{\rho}^* = \mathsf{\Psi}\bm{\alpha}^*$. 

In this setting, the compressed sensing theory shows that if the number of measurements satisfies
\begin{equation}
\label{cs5}
M \geq DNK\mu^{2}\left(\mathsf{\Phi}^{\star},\mathsf{\Psi}\right)\ln^{4}N,
\end{equation}
for a universal constant $D$, then $\bm{\alpha}$ is recovered accurately by solving the $\rm BP$ problem (\ref{cs3}) \cite{rauhut10}. In relation (\ref{cs5}), $\mu\left(\mathsf{\Phi}^{\star},\mathsf{\Psi}\right)$ is the mutual coherence between the sparsity and sensing bases. It is defined as the maximum projection, in absolute value, between the sparsity basis vectors $\bm{\varphi}_{i}$, $1 \leq i \leq N$, and sensing bases vectors  $\bm{\psi}_{i'}$, $1 \leq i' \leq N$ \cite{candes07, rauhut10}:
\begin{equation}
\label{cs6}
N^{-1/2} \leq \mu\left(\mathsf{\Phi}^{\star},\mathsf{\Psi}\right) =  \max_{1\leq i,i'\leq N}\vert \bm{\varphi}_{i} \bm{\cdot} \bm{\psi}_{i'} \vert \leq 1.
\end{equation}
\begin{figure}
\centering
\includegraphics[width=7cm, height=10cm]{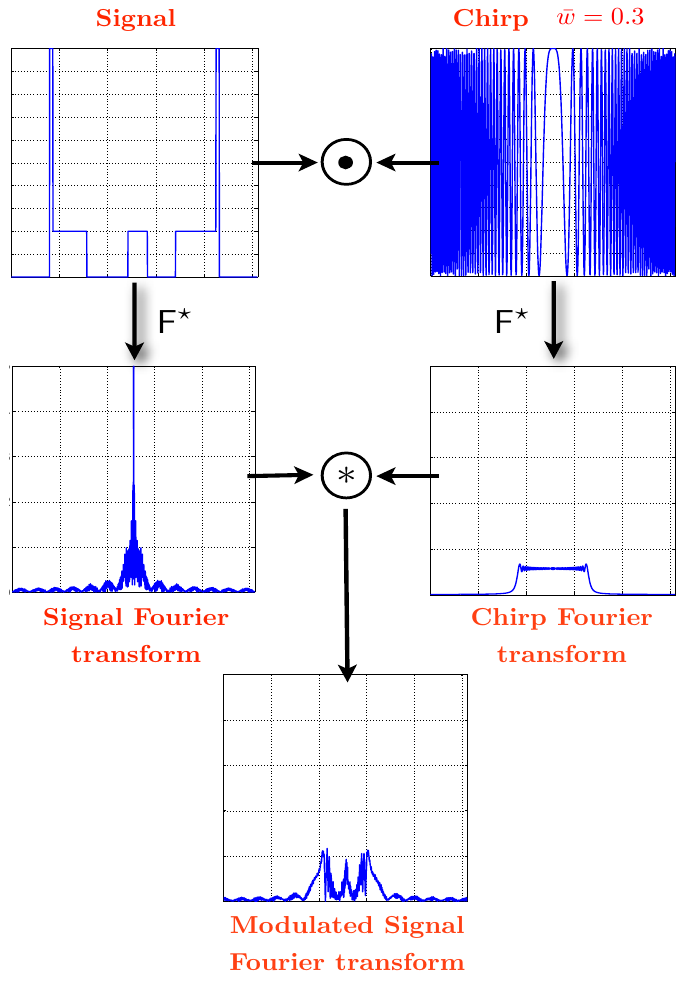}
\caption{\label{fig:ssp principle} Spread spectrum principle. A signal $\bm{\rho}$ (top left panel) is modulated by a linear chirp (top right panel). In $\bm{k}$-space, the modulation amounts to the convolution of the Fourier transform of the signal (middle left panel) with that of the chirp (middle right panel). The spectrum of the resulting signal (bottom panel) spreads out in $\bm{k}$-space.}
\vspace{-1mm}
\end{figure}

The mutual coherence $\mu\left(\mathsf{\Phi}^{\star},\mathsf{\Psi}\right)$ plays a crucial role in relation (\ref{cs5}). Indeed, the number of measurements $M^*$ needed to reconstruct $K$-sparse signals increases quadratically with its value: $M^*\propto\mu^{2}(\mathsf{\Phi}^{\star},\mathsf{\Psi})$. In the worst case where $\mu\left(\mathsf{\Phi}^{\star},\mathsf{\Psi}\right) = 1$, $M^*$ is of the order of the signal dimension $N$ and under-sampling is impossible. However, when the mutual coherence is at its minimum, $N^{-1/2}$, $M^*$ is reduced to the order of the sparsity level $K$. This result can intuitively be explained with a simple consideration on the spread of the energy of the sparsity basis vectors in the measurement domain $\mathsf{\Phi}$. In an incoherent orthonormal system, the absolute value of the scalar product between the sensing basis vector $\bm{\varphi}_{i}$ and the sparsity basis vector $\bm{\psi}_{i'}$ is small for all pairs of indices $1\leq i, i' \leq N$. As $\| \mathsf{\Phi}^{\star} \bm{\psi}_{i'} \|_2 = \| \bm{\psi}_{i'} \|_2$, the energy of the sparsity basis vector $\bm{\psi}_{i'}$ spreads equally over the sensing basis vectors $\bm{\varphi}_{i}$. Consequently, whatever index $i$ is selected to perform a measurement, one always gets information concerning all the sparsity basis vectors describing the original signal. The number of measurements needed for accurate recovery thus decreases.

%
%
% Spread spectrum and coherence
%
%
\subsection{Spread spectrum and coherence reduction}
\label{sub:spread spectrum and coherence}

\begin{table}
\caption{\label{tab:table Nmu} Values of $N_{\rm c} \, \mu_{\rm c}^2$ at different chirp rates $\bar{w}$}
\centering
\renewcommand{\arraystretch}{1.05}
\begin{tabular}{c|ccc}
\hline 
$N_{\rm c} \, \mu_{\rm c}^2$ & Dirac & Haar & Fourier \\
\hline\hline
$\bar{w} = 0$ & $1.00$ & $256$ & 256 \\
$\bar{w} = 0.1$ & $2.27$ & $43.5$ & 15.5 \\
$\bar{w} = 0.3$ & $2.97$ & $25.9$ & 6.11 \\
$\bar{w} = 0.5$ & $3.29$ & $22.4$ & 4.17 \\
\hline
\end{tabular}
\end{table}

The above considerations suggest that to reduce to the number of measurements needed for accurate recovery of MRI signals, one can try to modify the MRI acquisition procedure in such a way that the energy of the sparsity basis vectors spreads all over the $\bm{k}$-space. Following this idea, the s$_2$MRI technique introduces a linear chirp modulation of the signal of interest before random selection of $\bm{k}$-space coefficients (see Fig. \ref{fig:ssp principle}). Indeed, this modulation conserves the energy of the input signal and corresponds to a convolution that generically spreads the spectrum. 

To study theoretically the proposed acquisition scheme, we consider a simplified analog setting where the one-dimensional signal of interest is denoted by a complex-valued function $\rho(x)$ of the position $x \in \mathbb{R}$. This signal is limited on a finite field of view $L$. For the sake of simplicity of the following theoretical result only, we assume that the energy of this signal beyond the spatial frequency $B$ is negligible\footnote{In section \ref{sec:MR measurements}, we will introduce a setting that properly takes into account the fact that, in general, MRI signals are not band limited.}. We thus sample this signal on a discrete uniform grid of $N = 2LB$ points and represent the resulting discrete signal by a vector $\bm{\rho} \in \mathbb{C}^N$. The vector $\bm{\rho}$ is assumed to be $K$-sparse in an orthonormal basis $\mathsf{\Psi} \in \mathbb{C}^{N \times N}$. We also consider a one-dimensional linear chirp, with chirp rate $w \in \mathbb{R}$, that reads as a complex-valued function $c(x) = {\rm e}^{ {\rm i} \pi w x^2 }$. On the field of view $L$, this linear chirp is approximately band limited at its maximum instantaneous frequency $\vert w \vert L/2$. This band limit can be parametrized in terms of a discrete chirp rate $\bar{w} = w L^2/N$ and thus $\vert w \vert L/2 = \vert \bar{w} \vert B$. 

In this setting, the s$_2$MRI measurement model is given by equation (\ref{cs2}) with
\begin{eqnarray}
\mathsf{\Phi}^{\star} = \mathsf{F^{\star} CU} \in \mathbb{C}^{N_{\rm c} \times N}.
\end{eqnarray}
In the above equation, the matrix $\mathsf{U}$ represents an up-sampling operator needed to avoid aliasing of the modulated signal due to a lack of sampling resolution in a digital description of the originally analog problem. Indeed, the convolution in Fourier space induced by the analog chirp modulation implies that the band limit of the modulated signal is the sum of the individual band limits of the original signal and of the chirp $c$. Therefore, an up-sampled grid with at least $N_{\rm c}=(1+\vert\bar{w}\vert)N$ points needs to be considered and the modulated signal is correctly obtained by applying the chirp modulation on the signal after up-sampling on the $N_{\rm c}$ points grid. The up-sampling operator $\mathsf{U}$, implemented in Fourier space by zero padding, is thus of size $\mathbb{C}^{N_{c}\times N}$ and satisfies $\mathsf{U}^*\mathsf{U} = \mathsf{I} \in \mathbb{C}^{N \times N}$. The matrix $\mathsf{C} \in \mathbb{C}^{N_{c} \times N_{c}}$ is the diagonal matrix implementing the chirp modulation on the up-sampled grid and the matrix $\mathsf{F} = \left\{\bm{f}_i\right\}_{1 \leq i \leq N_{\rm c}}$ stands for the discrete Fourier basis on the same grid.

The s$_2$MRI sensing matrix $\mathsf{\Phi}$ is not orthogonal because of the presence of the matrix $\mathsf{U}$. Consequently, the recovery condition (\ref{cs5}) does not strictly hold. However, one can obtain a similar recovery condition \cite{puy11a}. In particular, if
\begin{eqnarray}
\label{eq:analog recovery}
M \geq D N_{\rm c} \, \mu_{\rm c}^2(\mathsf{\Phi}^{\star}, \mathsf{\Psi}) K \log^4(N),
\end{eqnarray}
for some universal constant $D$, then the vector $\bm{\rho}$ is accurately recovered with high probability by solving the $\rm BP$ problem (\ref{cs3}). This result shows that the number of measurements $M$ needed to reconstruct $K$-sparse signals is proportional to the product $N_{\rm c} \, \mu_{\rm c}^2(\mathsf{\Phi}^{\star},\mathsf{\Psi})$ rather than to $\mu_{\rm c}^2(\mathsf{\Phi}^{\star},\mathsf{\Psi})$ only. The s$_2$MRI technique is efficient only if this product decreases with the chirp rate $\bar{w}$. In other words, the number of measurements needed for accurate recovery of sparse signals decreases only for sparsity basis $\mathsf{\Psi}$ for which the mutual coherence $\mu_{\rm c}^2(\mathsf{\Phi}^{\star}, \mathsf{\Psi})$ decreases faster than $N_{\rm c}$ increases.

%
%
% Illustration
%
%
\subsection{Illustration}
\label{sub:illustration}

\begin{figure}
\centering
\includegraphics[width=9.2cm, keepaspectratio]{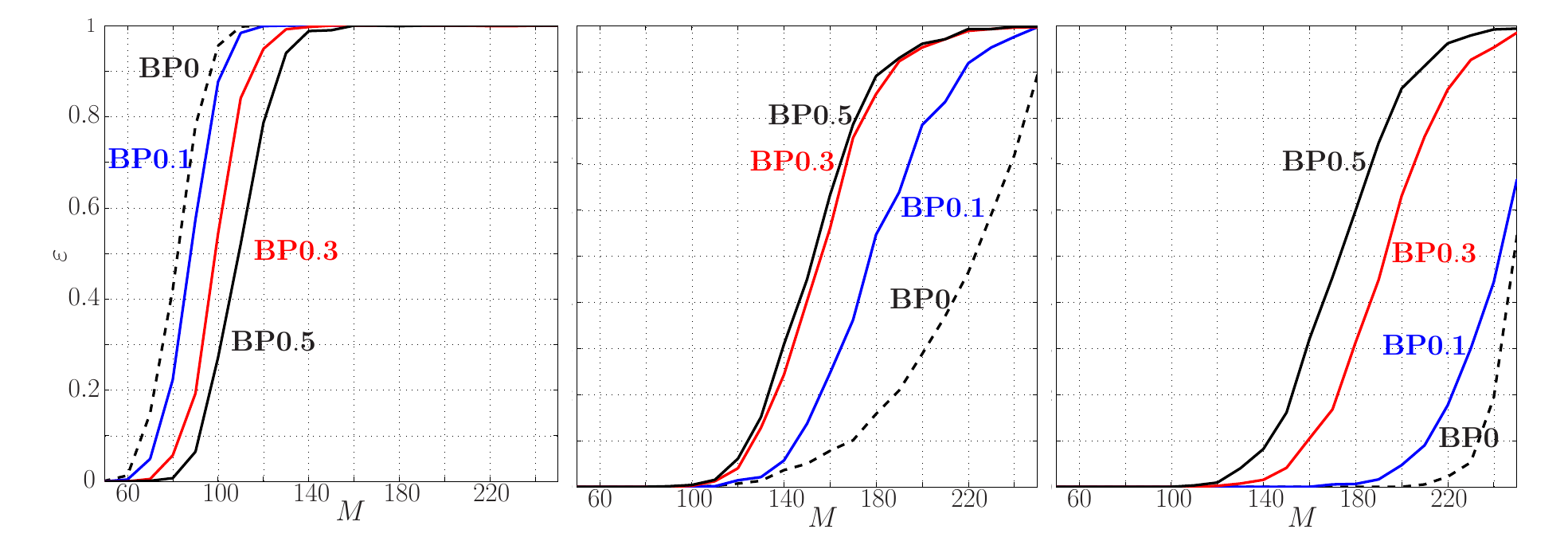}
\caption{\label{fig:theory ssp} Probability of recovery of the signal $\bm{\rho}$ as a function of the number of measurements $M$ for $\bar{w}=0$ (dotted black curve), $\bar{w}=0.1$ (continuous blue curve), $\bar{w}=0.3$ (continuous red curve), $\bar{w}=0.5$ (continuous black curve), and three sparsity bases: the Dirac basis (left); the Haar wavelet basis (middle); the Fourier basis (right).}
\end{figure}

We illustrate here the effect of the chirp modulation on the number of measurements needed for accurate recovery of $K$-sparse signals.

We consider a $1$-dimensional complex signal $\bm{\rho}$ of size $N = 256$ corresponding to one line of an MRI brain image. This signal is decomposed into three different sparsity bases $\mathsf{\Psi}$ and hard-thresholded at a fixed sparsity $K = 25$. The sparsity bases considered are the Dirac basis, the Haar wavelet basis and the Fourier basis. This signal is then probed according to relation (\ref{cs2}) with $\mathsf{\Phi^{\star}}=\mathsf{F^{\star}CU}$ and $\bar{w} \in \{0, 0.1, 0.3, 0.5\}$, and reconstructed from different numbers of measurements $M$ by solving the $\rm{BP}$ problem (\ref{cs3}). Each time, the probability of recovery\footnote{Perfect recovery is considered to occur if the $\ell_2$-norm between the original signal $\bm{\rho}$ and the reconstructed signal $\bm{\rho}^\star$ satisfies: $\|\bm{\rho}-\bm{\rho}^\star\|_2\leq10^{-3}\|\bm{\rho}\|_2$} of the signal is computed over $1000$ simulations. In this experiment, no noise is added to the measurements $\bm{y}$, and the indices $i$ of the selected Fourier basis vectors $\bm{f_i}$ are chosen uniformly at random from $\{0, \ldots, N_{\rm c}-1\}$.

The ${\rm BP}$ reconstructions in the presence of chirp modulation with chirp rate $\bar{w}$ are denoted by ${\rm BP}\bar{w}$. For each sparsity basis and chirp rate considered, the curves of the probability of recovery as a function of the number of measurements $M$ are reported in Fig. \ref{fig:theory ssp}. The corresponding values of the product $N_{\rm c} \, \mu_{\rm c}^2(\mathsf{\Phi}^{\star}, \mathsf{\Psi})$ are reported in Table \ref{tab:table Nmu}. 

For the Dirac basis, the number of measurements needed to reach a probability of recovery of $1$ slightly increases with the chirp rate, as suggested by the values of $N_{\rm c} \, \mu_{\rm c}^2(\mathsf{\Phi}^{\star}, \mathsf{\Psi})$ in Table \ref{tab:table Nmu} and relation (\ref{eq:analog recovery}). On the contrary, for the Haar and Fourier bases, the values of $N_{\rm c} \, \mu_{\rm c}^2(\mathsf{\Phi}^{\star}, \mathsf{\Psi})$ in the table predict a drastic improvement of the results with an increase of the chirp rate. This prediction is confirmed by the results presented in Fig. \ref{fig:theory ssp}. Note that for the Haar and Fourier bases, the value of the product $N_{\rm c} \, \mu_{\rm c}^2(\mathsf{\Phi}^{\star}, \mathsf{\Psi})$ decreases much less between $\bar{w} = 0.3$ and $\bar{w} = 0.5$ than between $\bar{w} = 0.1$ and $\bar{w} = 0.3$, suggesting a smaller improvement in the number of measurements needed for accurate recovery. This is also in line with the curves of the probability of recovery in Fig. \ref{fig:theory ssp}. In summary, as predicted by the theory, the effect of the s$_2$MRI technique depends on the sparsity basis. Note that the decrease of the performance for the Dirac basis (optimally incoherent only at $\bar{w} = 0$) is negligible compared to the improvement obtained for the two other bases. Note also that MRI signals are usually sparse in wavelet bases \cite{lustig07}. The results obtained with the Haar wavelet basis therefore suggest strong efficiency of the technique in MRI.

One can wonder if the proposed encoding scheme can result in shift-variant reconstruction quality. Indeed, the phase variation at the center of the chirp is less rapid than at the limit of the field of view. Let us consider the case of a signal sparse in the Haar wavelet basis. This signal can, roughly, be separated in large scale and fine scale sparsity basis vectors. The region where the chirp is oscillating slowly is limited to a small part of the field of view. Consequently, large scale sparsity basis vectors are necessarily affected by the chirp modulation as they have a wide support in signal space. The energy of these basis vectors is thus spread in $\bm{k}$-space thus improving the reconstruction quality of the low scale structures of the signal. On the other hand, the fine scale sparsity basis vectors at the center of the field of view remains not significantly modulated. However, these vectors are already incoherent with the Fourier basis as their energy naturally spreads out in $\bm{k}$-space. The fine scale structure of the signal are well recovered in absence and presence of chirp modulation. In summary, almost no shift-variant reconstruction quality is to be expected. 

Let us acknowledge that the idea of convolving the $\bm k$-space to optimize the acquisition procedure in the context of compressed sensing can also be found in \cite{sebert08, liang09b, wang09}. In these works, the $\bm{k}$-space is convolved by a random Toeplitz matrices. We should also note that the spread spectrum technique can be related to the random convolution approach where the signal is convolved by a random sequence and under-sampled in \emph{real space} \cite{romberg09}. In our case, convolution and under-sampling occur in \emph{$\bm k$-space}. Finally, let us mention that in a discrete setting, i.e., in the absence of band limit extension, replacing the linear chirp modulation by a random modulation leads to a universal encoding strategy, i.e., the reconstruction quality does not depend on the sparsity basis \cite{puy11a}. The s$_2$MRI technique tries to emulate this universal encoding strategy. A universal compressed sensing strategy might as well be obtained by projecting the signal $\bm{\rho}$ onto random waveforms with Gaussian distributions \cite{candes06c, donoho06}. In the context of MRI, an encoding strategy based on this sensing scheme was recently proposed in \cite{haldar10}.

%
%
%  MR MEASUREMENTS
%
%
\section{Spread spectrum MRI (s$_2$MRI)}
\label{sec:MR measurements}

\begin{figure}
\centering
\includegraphics[width=8.8cm, keepaspectratio]{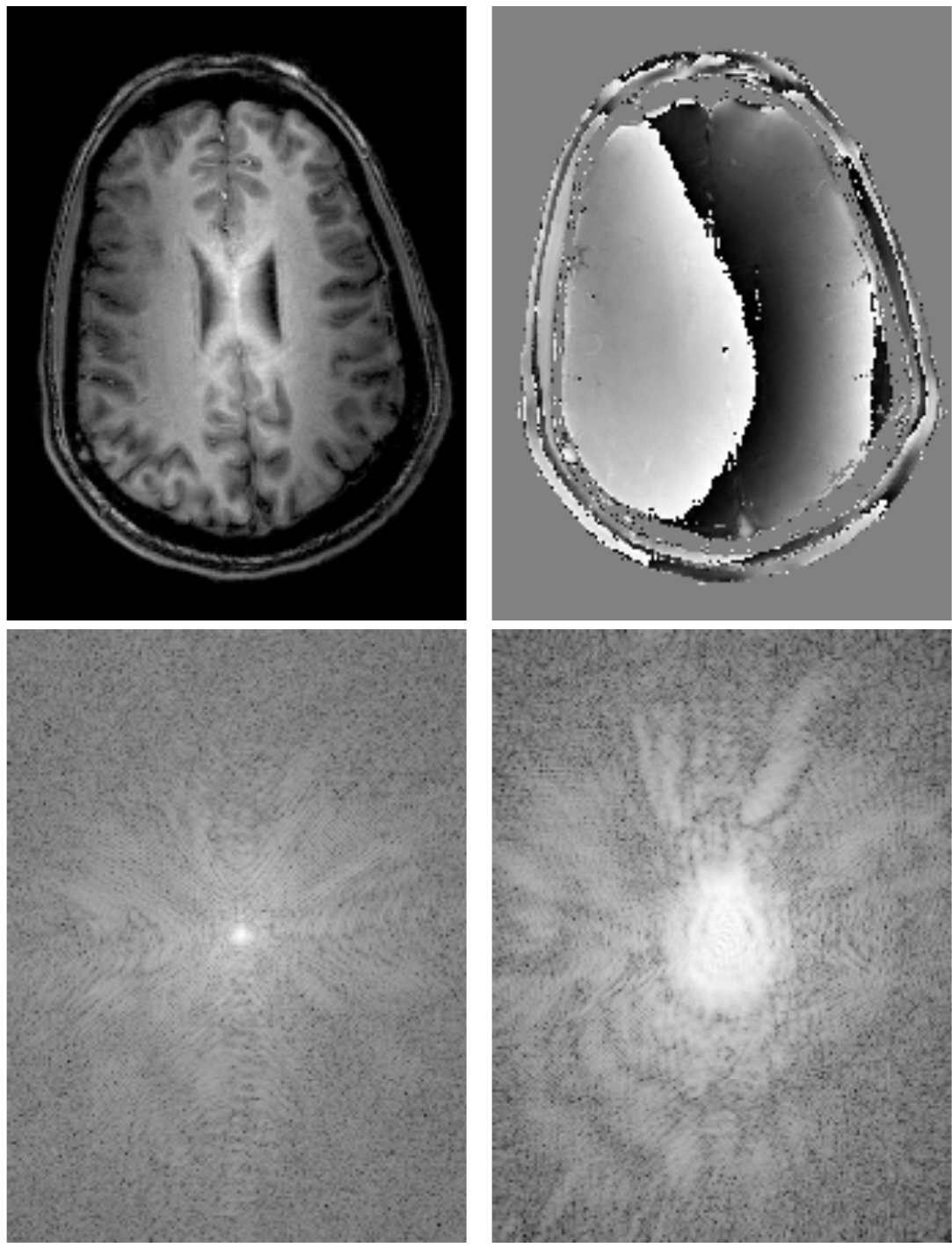}
\caption{\label{fig:original image}Original brain image at $1\,\rm{mm}$ of resolution. The first panel represents its absolute value. The second panel represents its phase mapped between $-\pi$ and $\pi$. The last two panels represent the logarithm of the amplitude of its Fourier transform before (third panel) and after (fourth panel) chirp modulation with chirp rate $\bar{w} = 0.3$. Dark and light regions respectively indicate low and high intensities.}
\end{figure}
%

%
%
% Quadratic phase profiles
%
%
\subsection{Quadratic phase profiles}
\label{sub:Quadratic phase profiles}

The spread spectrum principle was explained in the previous section in a simplified analog framework. Here, we move to a setting encompassing realistic analog MR signals which are limited in field of view and are consequently \emph{not} band limited.

Standard MR measurements take the form of  $\bm{k}$-space coefficients, with $\bm{k}=(k_x, k_y, k_z)$, of the original three-dimensional image probed representing the tissue magnetization as a function of the position inside a given field of view. This image is complex-valued due to magnetic field inhomogeneities \cite{haake99}. We consequently denote it by a complex-valued function $\rho(\bm{x})$ of the position $\bm{x}\in\mathbb{R}^{3}$, with components $(x,y,z)$ in image space. $(x,y)$ conventionally corresponds to the phase encoding directions, and $z$ corresponds to the readout direction. The field of view is $L=L_x\times L_y\times L_z$. We also consider quadratic phase profiles represented by a linear chirp in the phase encoding directions
\begin{equation}
\label{mri0}
c\left(\bm{x}\right) = {\rm e}^{ {\rm i} \pi \left( w_x x^2 + w_y y^2\right) },
\end{equation}
with chirp rate $(w_x, w_y) \in \mathbb{R}^2$. This linear chirp is characterized by an instantaneous frequency $(w_xx, w_yy)$ at position $\bm{x}$. On the finite field of view $L$, it is therefore approximately a band-limited function with approximate band-limits $\left(\vert w_x\vert L_x/2,\vert w_y\vert L_y/2\right)$ in the phase encoding directions.

In this setting, MR measurements at spatial frequency $\bm{k}\in\mathbb{R}^{3}$ take the general form
\begin{equation}
\label{mri1}
\nu\left(\bm{k}\right) = \int_{\mathbb{R}^{3}} \rho\left(\bm{x}\right) c \left(\bm{x}\right) {\rm e}^{-2{\rm i}\pi \bm{k}\bm{\cdot}\bm{x}} \: {\rm d}^{3}\bm{x}.
\end{equation}
In other words, the measurement $\nu$ corresponds to the coefficient at spatial frequency $\bm{k}$ of a signal obtained as the product of the original image $\rho(\bm{x})$ with the linear chirp modulation $c(\bm{x})$. In the absence of modulation ($w_x=w_y=0$), the measurements simply reduce to standard $\bm{k}$-space measurements. In the presence of quadratic phase profiles, the modulation amounts to the convolution of the Fourier transform of the chirp with that of the original image.

%
%
% k-space under-sampling and inverse problem
%
%
\subsection{Under-sampling in $\bm{k}$-space and the inverse problem}

\begin{figure}
\centering
\includegraphics[width=8.8cm, keepaspectratio]{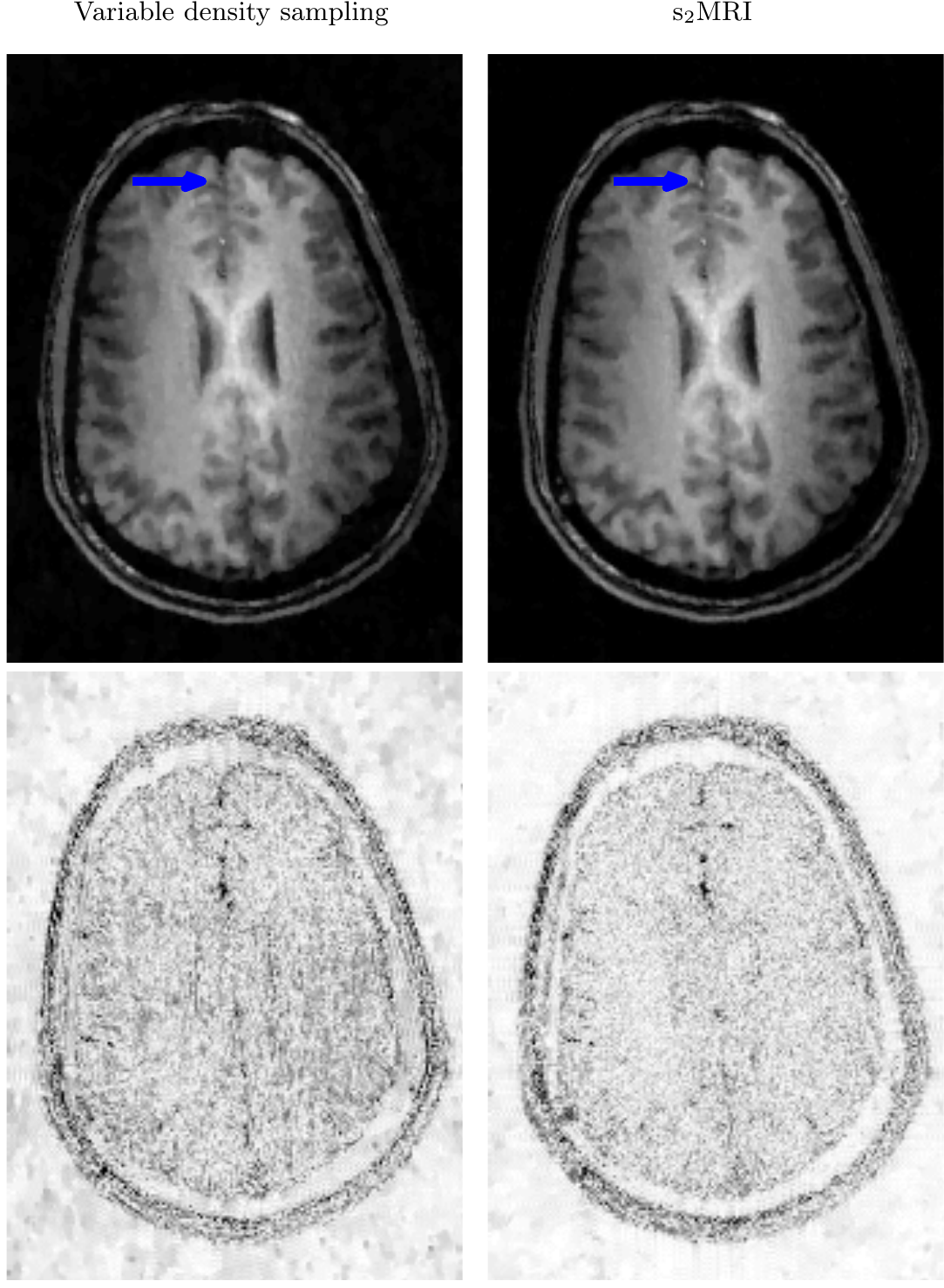}
\caption{\label{fig:simulated reconstruction 2D} Simulated reconstructions for an under-sampling of $20$ percent of $N$ and an input $\rm snr$ of $32$. The top row shows the reconstructions for the variable density sampling (left panel) and the s$_2$MRI (right panel) techniques, respectively. The bottom row shows the error images (difference between the original image and the reconstructions) in the absence (left panel) and presence (right panel) of chirp modulation. For a better visualization, the error images were scaled by a factor of $6$. The colormap for the error images goes from white to black, indicating low and high errors, respectively.}
\end{figure}
\begin{figure*}
\centering
\includegraphics[width=16.5cm]{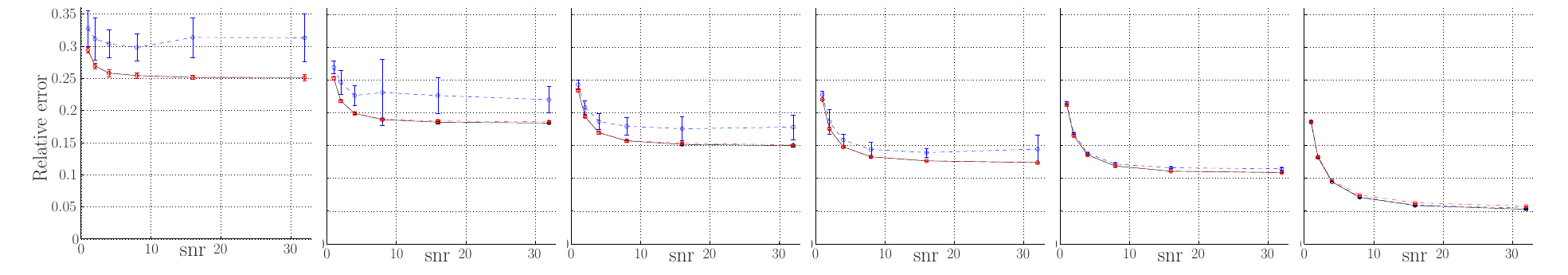}
\caption{\label{fig:curves error 2D} Relative reconstruction errors as functions of the input ${\rm snr}$ for the variable density sampling technique (dashed blue curve) and the s$_2$MRI technique with $\bar{w}=0.3$ (dot-dashed red curve) and $\bar{w} = \bar{w}_{\rm opt}$ (continuous black curve). The first to sixth panels show the curves for coverages of $5$, $10$, $15$, $20$, $25$ and $50$ percent of $N$ respectively. All curves represent the mean relative error over $30$ simulations, and the vertical lines represent the error at $1$ standard deviation.}
\vspace{-3mm}
\end{figure*}

Let us assume that we want to probe the signal $\rho$ at a resolution corresponding to a band-limit $B = (B_x, B_y, B_z)$. Note that it does not imply that the signal is band-limited and some energy may remain beyond $B$. At this resolution, the signal of interest is discretized on a discrete uniform grid of $N=N_x\times N_y\times N_z$ spatial frequencies $\bm{k}_{i}\in\mathbb{R}^{3}$, $1\leq i\leq N$, in $\bm{k}$-space, with $N_x=2L_xB_x$, $N_y=2L_yB_y$, and $N_z = 2L_zB_z$. In real space, this discretized signal may equivalently be described by its coefficients on a discrete uniform grid of $N$ points $\bm{x}_{i}$ with $1\leq i\leq N$. On this discrete grid, the linear chirp can be parametrized in terms of discrete chirp rates $(\bar{w}_x, \bar{w}_y)=\left(w_xL_x^2/N_x, w_yL_y^2/N_y\right)$, thus exhibiting approximate band-limits $(\vert \bar{w}_x \vert B_x, \vert \bar{w}_y \vert B_y, 0)$ on the finite field of view $L$.

We assume that the spatial frequencies probed span the $\bm{k}$-space up to the band-limits $B$. We also assume that these frequencies belong to the discrete grid of points $\bm{k}_{i}$, so that we can avoid any re-gridding operation \cite{haake99}. Due to the linear chirp modulation, the measurements can contain significant energy from $\bm{k}$-space coefficients beyond the band-limits $B$. Considering the estimated band-limits of the linear chirp, we choose to reconstruct the original signal $\rho$ on a high resolution grid of $N_{\rm c}=N_x(1+\vert\bar{w}_x\vert)\times N_y(1+\vert\bar{w}_y\vert)\times N_z$ points in order to prevent any aliasing in the reconstruction algorithm from these high spatial frequencies. The sampled signal on this high resolution grid is denoted by a vector $\bm{\rho} \in \mathbb{C}^{N_{\rm c}}$. The reconstructed signal is subsequently down-sampled at the desired resolution by keeping only its spatial frequencies belonging to the previously defined grid of size $N$.

The $\bm{k}$-space coverage provided by the $M$ spatial frequencies probed $\bm{k}_b$, with $1\leq b\leq M$, can be represented by a binary mask in $\bm{k}$-space, equal to $1$ for each spatial frequency probed and $0$ otherwise. The measurements may be denoted by a vector of $M$ $\bm{k}$-space coefficients $\bm{\nu}=\{\nu(\bm{k}_b)\}_{1\leq b\leq M}\in\mathbb{C}^{M}$, possibly affected by noise, which is denoted by the vector $\bm{n}=\{n_b\}_{1\leq b\leq M}\in\mathbb{C}^{M}$. In this setting, we consider an incomplete $\bm{k}$-space coverage ($M<N$), in order to accelerate the acquisition time in comparison with a complete $\bm{k}$-space coverage. In order to satisfy standard MRI constraints, an arbitrary under-sampling can be considered in the phase encoding directions while all spatial frequencies in the readout direction $k_z$ are probed. This form of under-sampling can directly be expressed in terms of an acceleration of the acquisition. Thus, the s$_2$MRI measurement model satisfies
\begin{equation}
\label{mri2}
\bm{\nu}=\mathsf{MF^{\star}CU}\bm{\rho}+\bm{n},
\end{equation}
where the matrix $\mathsf{MF^{\star}CU}\in\mathbb{C}^{M\times N_{\rm c}}$ encodes the complete linear relation between the signal and the measurements. The rectangular matrix $\mathsf{U}\in\mathbb{C}^{N_{\rm u}\times N_{\rm c}}$ represents an up-sampling operation, implemented in $\bm k$-space space by zero padding. This operation is needed to avoid any aliasing of the modulated signal due to a lack of sampling resolution in a discrete description of the originally continuous problem. The modulated signal is correctly obtained by applying the chirp modulation to the signal after up-sampling on the grid of $N_{\rm u} = N_x(1+2\vert\bar{w}_x\vert)\times N_y(1+2\vert\bar{w}_y\vert)\times N_z$ points by zero padding in the $k_x$ and $k_y$ directions. The matrix $\mathsf{C}\in\mathbb{C}^{N_{\rm u}\times N_{\rm u}}$ is the diagonal matrix implementing the chirp modulation on the up-sampled grid. The unitary matrix $\mathsf{F}\in\mathbb{C}^{N_{\rm u}\times N_{\rm u}}$ stands for the discrete Fourier basis on this high resolution grid. The matrix $\mathsf{M}\in\mathbb{R}^{M\times N_{\rm u}}$ is the rectangular binary matrix implementing the mask. It contains only one non-zero value on each line, at the index of the $\bm k$-space coefficient corresponding to each of the spatial frequencies probed.

Regarding the reconstruction of the signal $\bm{\rho}$, (\ref{mri2}) represents the measurement constraint. We take a statistical point of view and consider independent Gaussian noise on each measurement. Considering a candidate reconstruction $\bar{\bm{\rho}}$, the residual noise reads as $\bar{\bm{n}}=\bm{\nu}-\mathsf{MF^{\star}CU}\bar{\bm{\rho}}$. The noise level estimator, defined as twice the negative logarithm of the likelihood associated with $\bar{\bm{\rho}}$, reads as
\begin{equation}
\label{mri3}
\chi^{2}\left(\bar{\bm{\rho}},\bm{\nu}\right)=\sum_{b=1}^{M}\frac{\vert\bar{n}_{b}\vert^2}{[\sigma^{(n_b)}]^2},
\end{equation}
where the symbol $\vert\cdot\vert$ stands for the complex norm and $\sigma^{(n_b)}$ for the standard deviation of each of the real and imaginary parts of the noise component $n_b$. This noise level estimator follows a chi-square distribution with $2M$ degrees of freedom. Typically, this estimator should be minimized by a good candidate for the reconstruction. As explained in \cite{candes06b}, the measurement constraint on the reconstruction may be defined as a bound $\chi^{2}(\bar{\bm{\rho}},\bm{\nu})\leq\epsilon^{2}$, with $\epsilon^{2}$ corresponding to some large percentile of the chi-square distribution. In the remainder of the paper, we choose the 99$\rm th$ percentile.

%
%
% Variable density sampling
%
%
\subsection{Variable density sampling}
\label{sub:vds}

As discussed in \cite{lustig07}, the method selecting the spatial frequencies $\bm{k}_b$, with $1 \leq b \leq M$, is critical to achieve good reconstruction. Their approach is based on the use of a variable density sampling method. The distributions, which are empirically identified to provide optimized reconstructions, are defined by a power law decaying function $f (\bm{k}) = (1- \vert \bm{k}\vert/\vert\bm{k}_{m}\vert)^p + \beta$ for some power $p > 0$ and real constant $\beta \in [-1, 1]$ ($\vert \bm{k} \vert^2 = k_x^2+k_y^2$ and $\bm{k}_{m}$ is the highest spatial frequency of the $\bm{k}$-space domain to probe). In the present work, the spatial frequencies $\bm{k}_b$ are selected using the method\footnote{Toolbox available at \url{http://www.stanford.edu/~mlustig/SparseMRI.html}} of \cite{lustig07}: the values $f(\bm{k}_i)$ are first thresholded to restrict them to the interval $[0, 1]$; $N$ independent binary random variables $\delta \left(\bm{k}_i\right)$ taking value $1$ with probability $f \left(\bm{k}_i\right)$ are then generated; and finally the mask $\mathsf{M}$ selecting the $\bm{k}$-space coefficients for which $\delta \left(\bm{k}_i\right) = 1$ is created. 

Let us highlight the importance of the value $\beta$ on the actual shape of the variable density profile. For a fixed power $p$, this constant is computed beforehand in order to ensure that the number of measurements is, on average, equal to its target value $M$. Given a fixed number $M$ of measurements, we denote $p_M$ the value of $p$ for which $\beta = 0$. For a small value of $p$ ($p<p_M$), one would intuitively predict that the measurements spread all over the $\bm{k}$-space domain with the density of points higher at the center of $\bm{k}$-space. In fact, for $p<p_M$, one has $\beta<0$ and a whole $\bm{k}$-space region at high frequency remains unprobed. On the contrary, we have $\beta>0$ for $p>p_M$ and the $\bm{k}$-space is probed with a non-zero probability at the edges. Consequently, $p=p_M$ is the first power that ensures that the entire $\bm{k}$-space domain is probed with a non-zero probability. Also note that for $p>p_M$, the center of the $\bm{k}$-space is fully sampled, as $\beta>0$, and that the size of the fully sampled region increases when $p$ increases. 

In practice, the choice of the power $p$ of the variable density profile is difficult and should be adapted with the number of measurements in order to obtain the best reconstruction qualities. In the absence of chirp modulation, we performed reconstructions for several values of $p$ and noticed that $p=p_M$ leads to the best performance\footnote{In practice, $p_M$ is determined by an iterative process. Starting from $p = 0$, we increase its value by $0.5$ until we obtain $\beta \geq 0$ for the number of measurements considered.}. In the presence of a linear chirp modulation, the power spectrum of the original signal is spread and flattened. However, in the range of the chirp rates studied, the power spectrum of the modulated signal remains peaked at the origin (Fig. \ref{fig:original image}). We therefore also apply a variable density sampling and choose the power $p$ to be $p_M$. Empirically, this value also leads to the best reconstruction qualities. Note that choosing $p = p_M$ seems reasonable, as, given the spread of the information in the phase encoding directions, one wants to distribute the measurements over the entire sampling region of size $B$ and also limit the size of the fully sampled region at the $\bm{k}$-space center. 

Let us acknowledge that recent theoretical results obtained in \cite{puy11c} support the choice of such profiles in MRI.

%
%
% Simulation
%
%
\subsection{Numerical simulations}
\label{sub:simulations}

%
%
% Simulation protocol
%
%
\subsubsection{Simulation protocol}

To test the proposed technique, a real brain image is acquired on a $7\rm{T}$ short bore, actively shielded, MR scanner (Siemens, Erlangen, Germany). The subjects provided written informed consent prior to the imaging session, according to the guidelines of the local ethics committee. The parameters of the acquisition are as follows: $L_x \times  L_y \times L_z = 224 \times 168 \times 208 \, {\rm mm}^3$ with a resolution of $0.5 \times 0.5 \times 4 \, {\rm mm}^3$. The matrix size is thus $N_x \times  N_y \times N_z = 448 \times 336 \times 52$. The standard clinical ${\rm MPRAGE}$ sequence is used with echo time $\rm{TE} = 3.53\,\rm{ms}$, inversion time $\rm{TI} = 1.3 \,\rm{s}$, repetition time $\rm{TR} = 3.5\,\rm{s}$, and bandwidth $\rm{BW} = 200\,\rm{Hz}$. Note that, for the sake of simplicity, we restrict our analyses to one two-dimensional $z$-slice of the original three-dimensional acquisition with an under-sampling in both phase encoding directions $k_x$ and $k_y$. Also note that the image used is complex-valued.

In order to model an analog acquisition scheme, the original image at a resolution of $0.5 \, \rm{mm}$ is used to compute the measurements but the reconstruction is performed at a resolution of $1 \, \rm{mm}$. The reconstructed images are compared to the image obtained with a full acquisition at $1 \, \rm{mm}$ of resolution (Fig. \ref{fig:original image}).

The parameters of our analyses are as follows. Firstly, acquisitions are considered for various numbers $M$ of complex measurements corresponding to coverages of $5$, $10$, $15$, $20$, $25$ and $50$ percent of $N$. Secondly, instrumental noise is also added to the measurements as independent and identically distributed zero-mean Gaussian noise. The corresponding standard deviation $\sigma$ is identical for all the frequencies probed and we consider values of input ${\rm snr}$\footnote{The ${\rm snr}$ is defined as the ratio between the mean value of the complex magnitude of the original signal and the standard deviation of the noise $\sigma$.} of $2^j$, with $1\leq j \leq 6$. Thirdly, the chirp modulation studied has the same chirp rate $\bar{w}$ in both phase encoding directions, i.e., $\bar{w}=\bar{w}_x=\bar{w}_y$, with values in the range $[0, 0.5]$. Fourthly, the signals are reconstructed by solving the ${\rm BP}$ problem where the Total Variation (${\rm TV}$) norm of the signal $\bm \rho$ is substituted for the $\ell_1$ norm\footnote{The ${\rm TV}$ norm of a signal is defined as the $\ell_{1}$ norm of the magnitude of its gradient \cite{candes06a, rudin92}. \ Note that the recovery condition (\ref{eq:analog recovery}) does not hold with this norm. However, one can notice that the ${\rm TV}$ norm of a signal $\bm{\rho}$ is very similar to the $\ell_1$ norm of its decomposition in the Haar wavelet basis. In the light of the preliminary results of Section \ref{sub:illustration}, one can thus hope to obtain an improvement of the reconstruction quality in presence of chirp modulation.}. This problem is solved thanks the Douglas Rachford algorithm \cite{combettes11, fadili09}. Note that the ${\rm TV}$ norm in combination with wavelet sparsity basis is, for example, used for reconstructing MRI signals from under-sampled $\bm{k}$-space in \cite{lustig07, liang09a}, or \cite{kern11}. For our problem, we tested several multiscale representations such as Daubechies wavelets, steerable wavelets, or curvelets, but the best reconstructions were obtained with the ${\rm TV}$ norm. Finally, for each value of $M$ and input ${\rm snr}$ considered, $30$ simulations are generated with independent noise and mask realizations, and the relative reconstruction errors $\|\bm{\rho}-\bm{\rho}^\star\|_2/\|\bm{\rho}\|_2$ are computed for the variable density sampling and s$_2$MRI techniques. For each value of $M$ and input ${\rm snr}$, the discrete chirp rate $\bar{w}_{\rm opt}$ that gives the smallest relative error on average over the $30$ simulations is recorded\footnote{s$_2$MRI toolbox available at \url{http://lts2www.epfl.ch/people/gilles}.}.

%
%
% Simulation results
%
%
\subsubsection{Simulation results}

The magnitudes of the reconstructed images obtained with the s$_2$MRI and the variable density sampling techniques for an acceleration factor of $5$ and an input $\rm snr$ of $32$ are presented in Fig. \ref{fig:simulated reconstruction 2D}, along with the corresponding error images (magnitudes of the complex-valued differences between the original image and reconstructed images). The relative errors of the reconstructions as functions of the input ${\rm snr}$ for the six coverages considered and for both methods are reported in Fig. \ref{fig:curves error 2D}. 

For acceleration factors larger than $2$, the s$_2$MRI technique with $\bar{w} = 0.3$ provides better reconstruction than the variable density sampling technique with an improvement up to $0.05$ of the relative error. Indeed, the relative error is, on average, lower in the presence of the chirp modulation. The corresponding standard deviations are also much smaller, indicating that the s$_2$MRI technique is more stable\footnote{Some leftover variability for the variable density sampling technique might still be removed by increasing the number of simulations. However, this would not modify the results and comparison with s$_2$MRI.}. At an acceleration factor of $2$, the variable density sampling technique gives slightly better reconstructions than the s$_2$MRI technique with $\bar{w} = 0.3$. However, with $\bar{w} = \bar{w}_{\rm opt}=0.1$, the s$_2$MRI technique provides relative errors similar to those obtained with the variable density sampling technique. These results suggest to reduce the chirp rate as the number of measurements increases. This is coherent with the fact that modulation is not needed in the limit of no under-sampling.

When comparing the magnitudes of reconstructed images in Fig. \ref{fig:simulated reconstruction 2D}, differences between both methods do not appear obvious. However, one can notice that fine details are recovered better with the s$_2$MRI technique: the vessel (white spot) indicated by a blue arrow appears in the s$_2$MRI reconstruction, but not in the variable density sampling reconstruction. The error images bring more information, and one can notice that the errors are smaller in the presence of chirp modulation. In particular, the low scale structures, rendered incoherent with the chirp modulation, are better recovered.

%
%
% REAL ACQUISITION
%
%
\section{s$_2$MRI implementation and experimental results}
\label{sec:Real acquisitions}

%
%
% Implementation
%
%
\subsection{Implementation}
The s$_2$MRI technique is tested on the $7\rm{T}$ scanner described in Section \ref{sub:simulations} with $3\rm{D}$ acquisitions of a phantom and a human brain. For the brain experiment, the subjects provided written informed consent prior to the imaging session, according to the guidelines of the local ethics committee.

The chirp modulation is implemented through the use of a second order shim coil $x^2-y^2$. In our implementation, the chirp rate varies linearly with the readout time $t$ (or equivalently $k_z$) and is proportional to the intensity of the quadratic magnetic field $\kappa$: $w(t) = w_x(t) = - w_y(t) = \gamma\kappa t / \pi $, where $\gamma$ is the gyromagnetic factor. The maximum chirp rate $w^{\rm max}$ is reached at $t = {\rm TE} + \Delta t/2$, the mean chirp rate $w^{\rm mean}$ at $t = {\rm TE}$, and the minimum chirp rate $w^{\rm min}$ at $t = {\rm TE} - \Delta t/2$ ($\Delta t$ is the readout duration). This chirp modulation can be introduced in the measurement model by modifying (\ref{mri2}) as follows:
\begin{equation}
\label{eq:real modulation}
\bm{\nu}=\mathsf{MF}^{\star}_{xy}\mathsf{CF}^{\star}_{z}\mathsf{U}\bm{\rho}+\bm{n}.
\end{equation}
In the above equation, $\mathsf{F}^{\star}_{z} \in \mathbb{C}^{N_{\rm u} \times N_{\rm u}}$ and $\mathsf{F}^{\star}_{xy} \in \mathbb{C}^{N_{\rm u} \times N_{\rm u}}$ implement the Fourier transform along the $z$-direction and the $xy$-directions, respectively. The matrix $\mathsf{C} \in \mathbb{C}^{N_{\rm u} \times N_{\rm u}}$ implements the chirp modulation. Signals are thus reconstructed on a grid of $N_{\rm c}=N_x(1+\vert\bar{w}_x^{\rm max}\vert)\times N_y(1+\vert\bar{w}_y^{\rm max}\vert)\times N_z$ points, and $N_{\rm u}=N_x(1+2\vert\bar{w}_x^{\rm max}\vert)\times N_y(1+2\vert\bar{w}_y^{\rm max}\vert)\times N_z$.

This modulation can be decomposed as a quadratic phase modulation in the $(x, y)$ planes with chirp rate $w^{\rm mean}$, combined with a linear phase modulation in $k_z$. This linear phase modulation produces shifts of the original signal by an amount proportional to $\kappa (x^2 - y^2)$ along the $z$ direction. The chirp modulation used is thus not ideal, as it creates distortions of the original object and complicates the measurement matrix. However, the energy of sparsity basis vectors is still spread by the main chirp modulation with chirp rate $w^{\rm mean}$, and the previous conclusions based on theory and simulations should still hold. This will be confirmed by the numerical experiments of Section \ref{sub:numerical experiments}. Note also that the reconstructed images are free of any distortion as the complete effect of the modulation is modeled in (\ref{eq:real modulation}). Let us remark that these distortions might be avoided with the use of RF pulses or dedicated coils applied only during phase encoding.

As in Section \ref{sub:simulations}, the s$_2$MRI technique is compared to the variable density sampling method with $p = p_M$. Full acquisitions ($M/N = 1$) are performed both in the absence and in the presence of the chirp modulation. The number of phase encodings $(k_x, k_y)$ is then reduced retrospectively by applying a mask on the complete data.

\begin{figure}
\centering
\includegraphics[width=8.9cm, keepaspectratio]{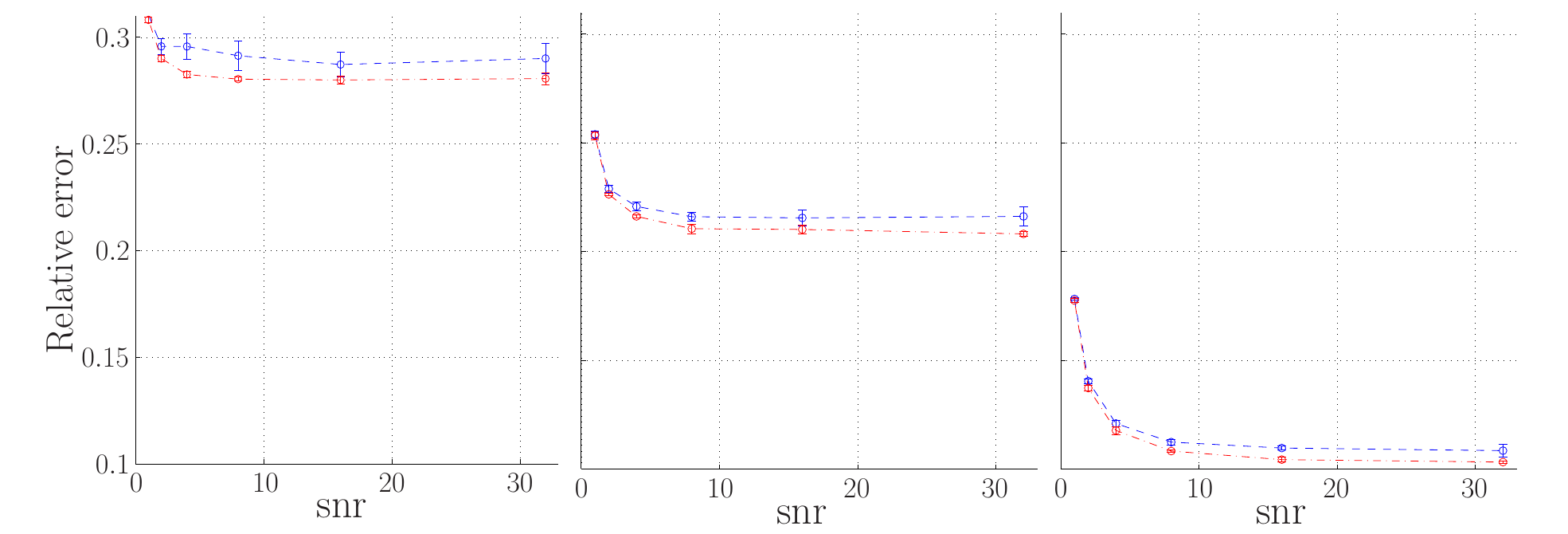}
\caption{\label{fig:curves error 3D} Relative reconstruction errors as functions of the input ${\rm snr}$ for the variable density sampling technique (dashed blue curve) and the s$_2$MRI technique (dot-dashed red curve) with varying chirp modulation. From left to right and top to bottom, the panels show the curves for coverages of $15$, $25$, $50$ percent of $N$ respectively. All curves represent the mean relative error over $5$ simulations, and the vertical lines represent the error at $1$ standard deviation.}
\vspace{-3mm}
\end{figure}

The noise level is evaluated directly on the under-sampled data available for reconstruction. For each $(k_x, k_y)$ pair measured, all the frequencies $k_z$ are probed so that the signal is available as a function of $z$. The level of the noise is estimated on probed pairs $(k_x, k_y)$ at positions $z$ that do not contain any signal. This noise level is identical in the presence and absence of chirp modulation.

%
%
% Simulations
%
%
\subsection{Numerical validation}
\label{sub:numerical experiments}

In this section, we perform simulations using the acquisition scheme described above to confirm that, even though the implementation of the chirp modulation is not ideal, the reconstruction quality is still enhanced with the s$_2$MRI technique.

\begin{figure}
\centering
\includegraphics[width=8.8cm, keepaspectratio]{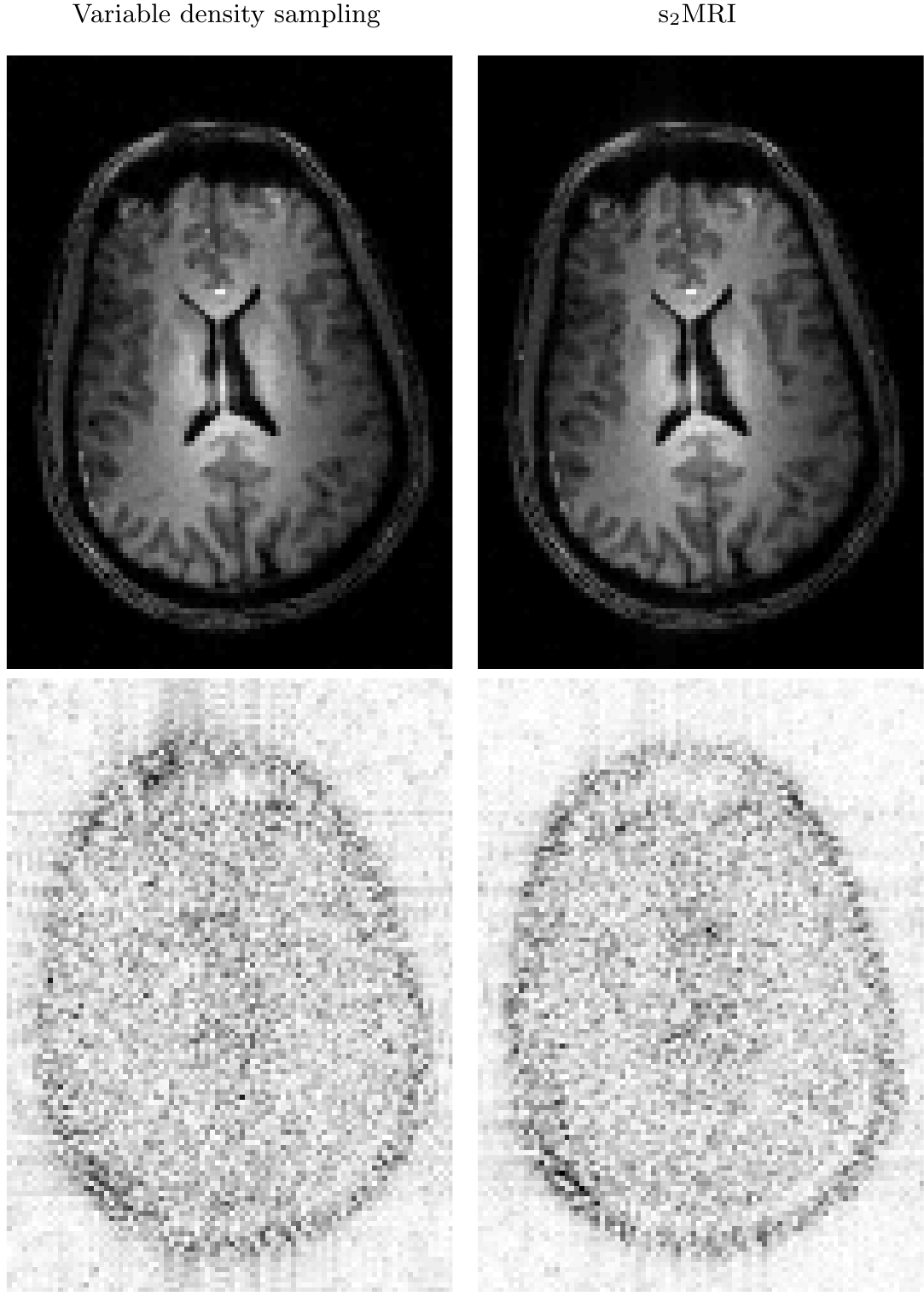}
\caption{\label{fig:simulated reconstruction 3D} Simulated reconstructions with varying chirp modulation for an under-sampling of $50$ percent of $N$ and an input $\rm snr$ of $32$. The top row shows the reconstructions of an axial slice for the variable density sampling (left panel) and the s$_2$MRI (right panel) techniques, respectively. The bottom row shows the error images (difference between the original image and the reconstructions) in the absence (left panel) and presence (right panel) of chirp modulation. For a better visualization, the error images were scaled by a factor of $8$. The colormap for the error images goes from white to black, indicating low and high errors, respectively.}
\vspace{-3mm}
\end{figure}

For this numerical experiment, a brain volume was acquired using the standard clinical MPRAGE sequence on a field of view of $L_x = 243\,\rm{mm}$, $L_y = 176\,\rm{mm}$ and $L_z = 256\,\rm{mm}$, with a resolution of $1\,\rm{mm}$ in each direction ($N_x = 243$, $N_y = 176$, $N_z = 256$). The echo time is $\rm{TE} = 4.59\,\rm{ms}$, the inversion time $\rm{TI} = 1.5\,\rm{s}$, the repetition time $\rm{TR} = 3.5\,\rm{s}$, the bandwidth $\rm{BW} = 250\,\rm{Hz}$. As in Section \ref{sub:simulations}, in order to model an analog acquisition scheme, the original image at a resolution of $1 \, \rm{mm}$ is used to compute the measurements, but the reconstruction is performed at a resolution of $2 \, \rm{mm}$. The reconstructed images are compared to the image obtained with a full acquisition at $2 \, \rm{mm}$ of resolution. 

The parameters of our experiment are as follows. Firstly, acquisitions are considered for various numbers $M$ of complex measurements corresponding to coverages of $15$, $25$, $50$ percent of $N$. Secondly, instrumental noise is also added to the measurements as independent and identically distributed zero-mean Gaussian noise. The corresponding standard deviation $\sigma$ is identical for all the frequencies probed and we consider values of input ${\rm snr}$ of $2^j$, with $1\leq j \leq 6$. Thirdly, the simulated chirp modulation has a chirp rate varying linearly with $k_z$: $\vert \bar{w}_x \vert \in [0.12, 0.30]$ and $\vert \bar{w}_y \vert \in [0.09, 0.22]$. These values for the discrete chirp rate correspond to those used during the real experiment performed hereafter at $1 \, \rm{mm}$ of resolution. However, relatively to the band-limit at $2 \, \rm{mm}$ of resolution, the spectrum is naturally more spread than relatively to the band-limit at $1 \rm{mm}$ of resolution. Therefore, for a reconstruction at $2 \, \rm{mm}$ of resolution, the spectrum does not need to be spread as much as for a reconstruction at $1 \, \rm{mm}$ of resolution. We thus divided the values of the chirp rate by $2$ in both dimensions. Fourthly, the signals are reconstructed by solving the ${\rm BP}$ problem where the ${\rm TV}$ norm of the signal $\bm \rho$ is substituted for the $\ell_1$ norm. Finally, for each value of $M$ and input ${\rm snr}$ considered, $5$ simulations are generated with independent noise and mask realizations, and the relative reconstruction errors are computed for the variable density sampling and s$_2$MRI techniques.

The relative errors of the reconstructions as functions of the input ${\rm snr}$ for the three coverages considered and for both methods are reported in Fig. \ref{fig:curves error 3D}. The magnitudes of a reconstructed axial slice obtained with the s$_2$MRI and the variable density sampling techniques for an acceleration factor of $5$ and an input $\rm snr$ of $32$ are presented in Fig. \ref{fig:simulated reconstruction 3D}, along with the corresponding error images. Conclusions of Section \ref{sub:simulations} still hold with this acquisition scheme. The relative error of reconstruction is lower in the presence of the chirp for the three coverage considered. When comparing the error images in Fig. \ref{fig:simulated reconstruction 3D}, one can once more notice that the errors are smaller with the s$_2$MRI technique.

%
%
% Experiments
%
%
\subsection{Experiments}

\begin{figure*}
\centering
\includegraphics[width=18.2cm, keepaspectratio]{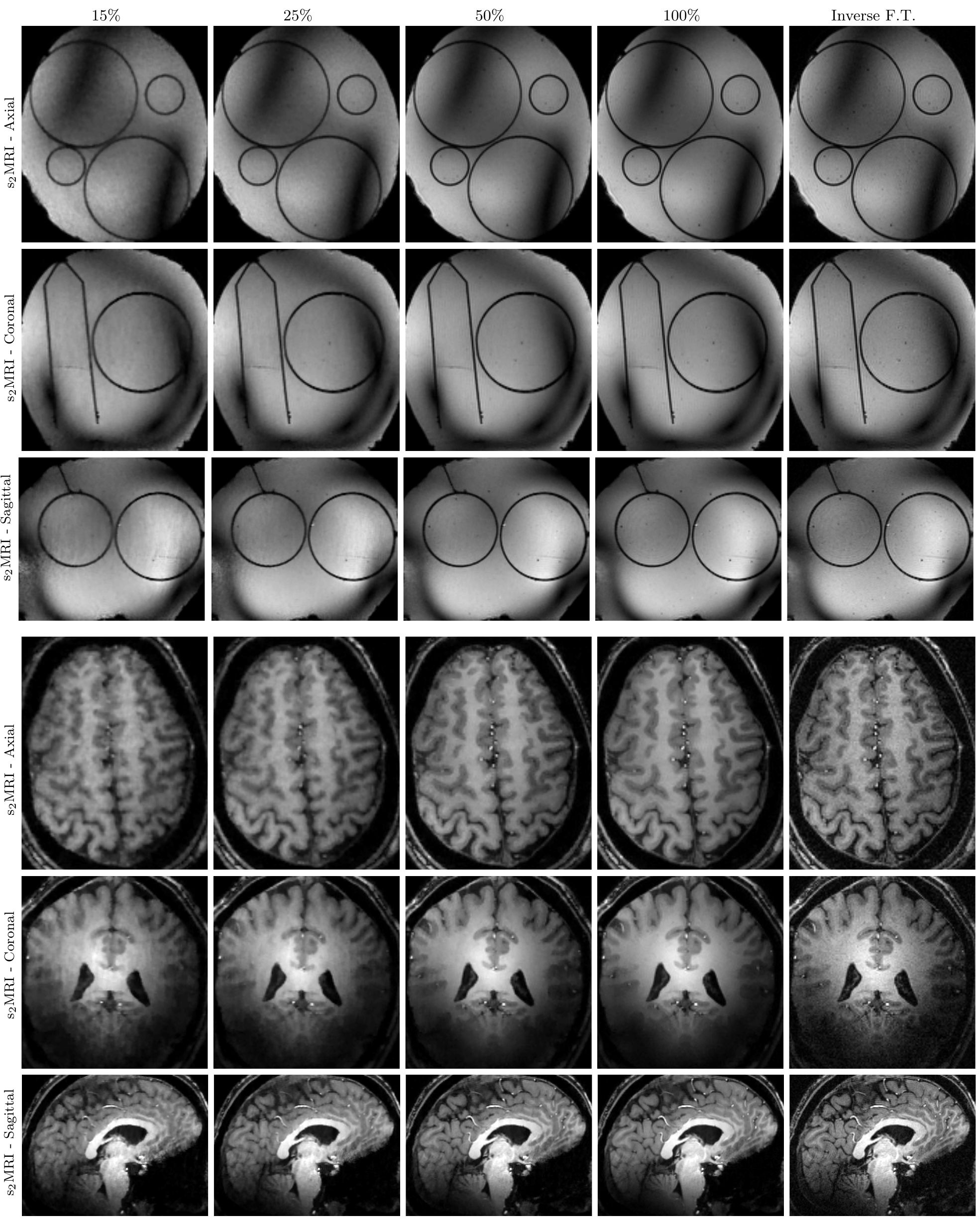}
\caption{\label{fig:experiments} Phantom and brain reconstructions from real experimental data with the s$_2$MRI technique. The first to fourth columns show the magnitude of the reconstructions from $15$, $25$, $50$, and $100$ percent of phase encodings respectively. The fifth column shows the reference images obtained by inverse Fourier transform (F.T.) of the fully sampled $\bm{k}$-space.}
\end{figure*}
\begin{figure*}
\centering
\includegraphics[width=18.2cm,keepaspectratio]{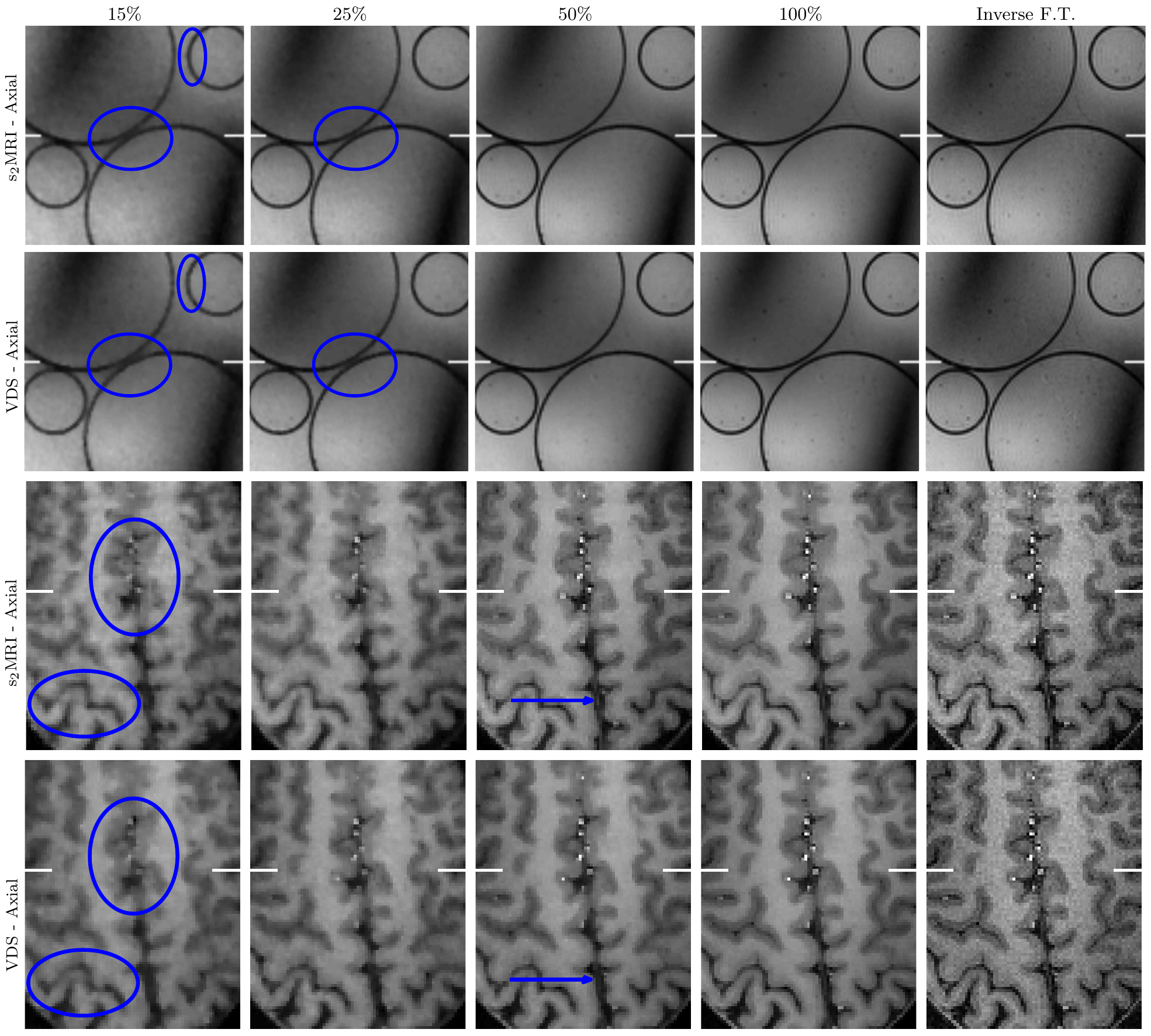}
\caption{\label{fig:comparisons images} Phantom and brain reconstructions from real experimental data with the s$_2$MRI technique (first and third rows respectively) and the variable density sampling (VDS) technique (second and fourth rows respectively). The first to fourth columns show the magnitude of the reconstructions from $15$, $25$, $50$, and $100$ percent of phase encodings respectively. The fifth column shows the reference images obtained by inverse Fourier transform (F.T.) of the fully sampled $\bm{k}$-space. The white lines on these images indicate the location of the spatial profiles presented in Fig. \ref{fig:comparisons profiles}}
\vspace{-4mm}
\end{figure*}
\begin{figure*}
\centering
\includegraphics[width=18.5cm, keepaspectratio]{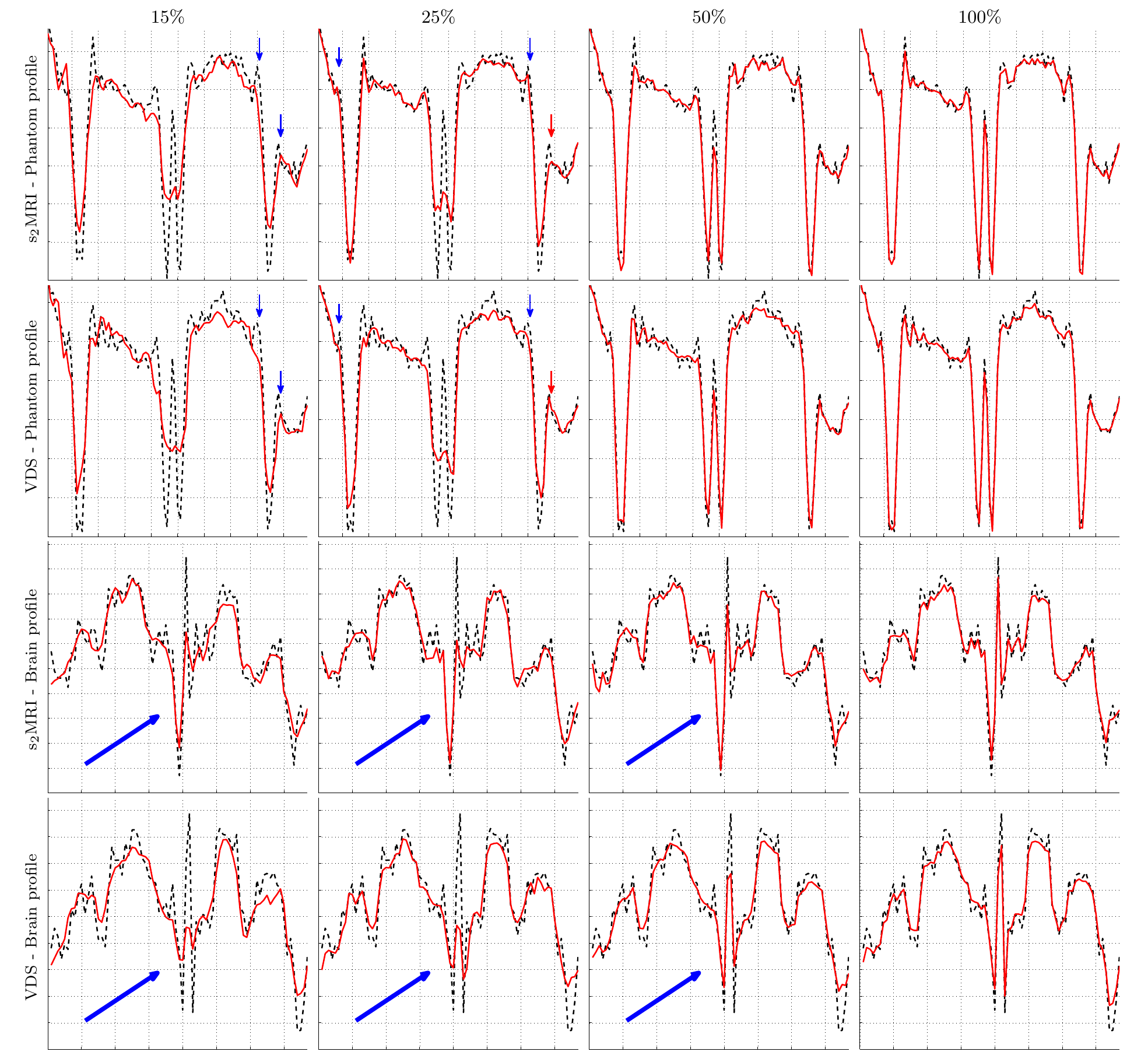}\\
\caption{\label{fig:comparisons profiles} Phantom and brain reconstruction profiles from real experimental data with the s$_2$MRI technique (first and third rows respectively) and the variable density sampling (VDS) technique (second and fourth rows respectively). The first to fourth columns show the profile magnitude of the reconstructions (continuous red curve) from $15$, $25$, $50$, and $100$ percent of phase encodings respectively as well as the reference profiles (dotted black curve) obtained by inverse Fourier transform (F.T.) of the fully sampled $\bm{k}$-space.}
\vspace{-4mm}
\end{figure*}

For the phantom experiment, a gradient echo sequence is used on a field of view of $L_x = L_y = L_z = 192\,\rm{mm}$ in the three directions with a resolution of $1\,\rm{mm}$ ($N_x = N_y = N_z = 192$). The echo time is $\rm{TE} = 6\,\rm{ms}$, the repetition time $\rm{TR} = 10\,\rm{ms}$, the bandwidth $\rm{BW} = 400\,\rm{Hz}$, and the quadratic magnetic field intensity $\kappa = 3 000\,\rm{\mu T/m^{2}}$. The discrete chirp rates satisfy $\vert \bar{w}_x \vert = \vert \bar{w}_y \vert \in [0.23, 0.36]$.

For the brain experiment, the standard clinical MPRAGE sequence is used on a field of view of $L_x = 243\,\rm{mm}$, $L_y = 176\,\rm{mm}$ and $L_z = 256\,\rm{mm}$, with a resolution of $1\,\rm{mm}$ in each direction ($N_x = 243$, $N_y = 176$, $N_z = 256$). The echo time is $\rm{TE} = 4.59\,\rm{ms}$, the inversion time $\rm{TI} = 1.5\,\rm{s}$, the repetition time $\rm{TR} = 3.5\,\rm{s}$, the bandwidth $\rm{BW} = 250\,\rm{Hz}$, and the quadratic magnetic field intensity $\kappa = 4 500\,\rm{\mu T/m^{2}}$. The discrete chirp rates satisfy $\vert \bar{w}_x \vert \in [0.25, 0.61]$ and $\vert \bar{w}_y \vert \in [0.18, 0.44]$.

Slices of the $3\rm{D}$ reconstructions obtained with the s$_2$MRI technique for $15$, $25$, $50$ and $100$ percent of phase encodings are presented in Fig. \ref{fig:experiments} for the phantom and brain acquisitions. The images obtained by inverse Fourier transform of the fully sampled $\bm{k}$-space (after correction of the distortions due to the small shifts in the $z$ direction) are also presented. Fig. \ref{fig:comparisons images} provides a comparison of the reconstructed images obtained with the s$_2$MRI and the variable density sampling techniques. In the aim of providing further insight on how each method preserves image resolution or, in other words, captures shape and magnitude of high frequency features, we also provide one-dimensional spatial profiles in Fig. \ref{fig:comparisons profiles}. The white lines in Fig. \ref{fig:comparisons images} indicate the location of these profiles.

Firstly, as one would expect, the visual reconstruction quality improves when the number of phase encodings $M$ increases. In the limit of a coverage of $100$ percent, we cannot identify any loss of details between the reconstructed images and the ones obtained by inverse Fourier transform. Moreover, the reconstructed image contains much less noise than the image obtained by inverse Fourier transform. Indeed, at a coverage of $100$ percent, the problem (\ref{cs3}) is essentially reduced to a denoising problem.

Secondly, as in Section \ref{sub:simulations}, the differences between both methods do not appear at first glance when comparing only the magnitudes of reconstructed images in Fig. \ref{fig:comparisons images}. Unfortunately here, the ground truth image is not accessible, so the corresponding error images cannot be displayed. However, a thorough visual inspection reveals that, for acceleration factors larger than $2$, some fine details are better recovered by our approach. On the phantom reconstructions for acceleration factors of $6.7$ and $4$ (coverages of $15$ and $25$ percent, respectively), the separation between the two biggest circles is more visible. The shapes of the circles are also more curved. On the brain reconstructions for an acceleration factor of $6.7$, the vessels at the center of image are still visible with our method but not with the variable density sampling technique. The cerebral cortex also appears sharper. For an acceleration factor of $2$, the thin layer separating the two hemispheres of the brain remains more visible with the s$_2$MRI technique.

Thirdly, the reconstructed spatial profiles of the phantom presented in Fig. \ref{fig:comparisons profiles} show that the shape and magnitude of high frequency features are, more often, slightly better recovered with the s$_2$MRI technique for acceleration factors larger than $4$ (blue and red arrows indicate features better reconstructed with s$_2$MRI and variable density sampling respectively). This improvement is much more significant on the brain data (see arrows), and holds for acceleration factors larger than $2$.

Finally, the distortions are correctly taken into account by the operator $\mathsf{C}$, as both fully sampled slices of the phantom with and without chirp are identical (see Fig. \ref{fig:comparisons images}). For the brain images, some differences remain due to small movements of the subject between both acquisitions. Let us remark that no fitting of the chirp parameters (center and rates) was performed to improve on the theoretical values. This highlights sufficient stability of the technique relative to parameter approximations.

%
%
% CONCLUSION
%
%
\section{Conclusion and discussion}
\label{sec:Conclusion}

We presented a spread spectrum technique (s$_2$MRI) designed to accelerate MR acquisitions by compressed sensing. It consists of pre-modulating the original image by a linear chirp, which results from the application of quadratic phase profiles, and then performing random $\bm k$-space under-sampling. Non-linear algorithms promoting signal sparsity are then used for image reconstruction.

In the context of compressed sensing theory, the effectiveness of the technique is supported by a decrease of coherence between the sensing and sparsity bases due to the pre-modulation. Simulations in a simplified analog setting confirm that the enhancement of the image reconstruction quality is linked to the evolution of the mutual coherence. The s$_2$MRI technique was compared with the state-of-the-art variable density sampling using realistic numerical simulations and real acquisitions. Simulation results shows that the s$_2$MRI technique performs slightly better than the variable density sampling technique in terms of relative reconstruction error for acceleration factors larger than $2$. The chirp modulation was also implemented on a $7$T scanner with the use of a second order shim coil. Simulations of this implementation confirms again the slight superiority of s$_2$MRI. Visual inspection of reconstructions obtained from real acquisition of phantom and \emph{in vivo} data also shows that this first (non-ideal) implementation provides slightly better reconstruction qualities than the variable density sampling method. The s$_2$MRI technique thus outperforms the variable density sampling technique in terms of all the criteria used for evaluation.

Regarding future evolutions of the s$_2$MRI technique, an implementation of the linear chirp modulation with the use of RF pulses or dedicated coils could simplify the measurement scheme by applying the chirp modulation only during phase encoding in order to avoid object distortions, and, in turn, further enhance the reconstruction quality. Moreover, fitting the effective chirp center and rates on the basis of the data could improve the measurement model and result in better reconstructions.

\begin{figure}
\centering
\includegraphics[width=8.8cm, keepaspectratio]{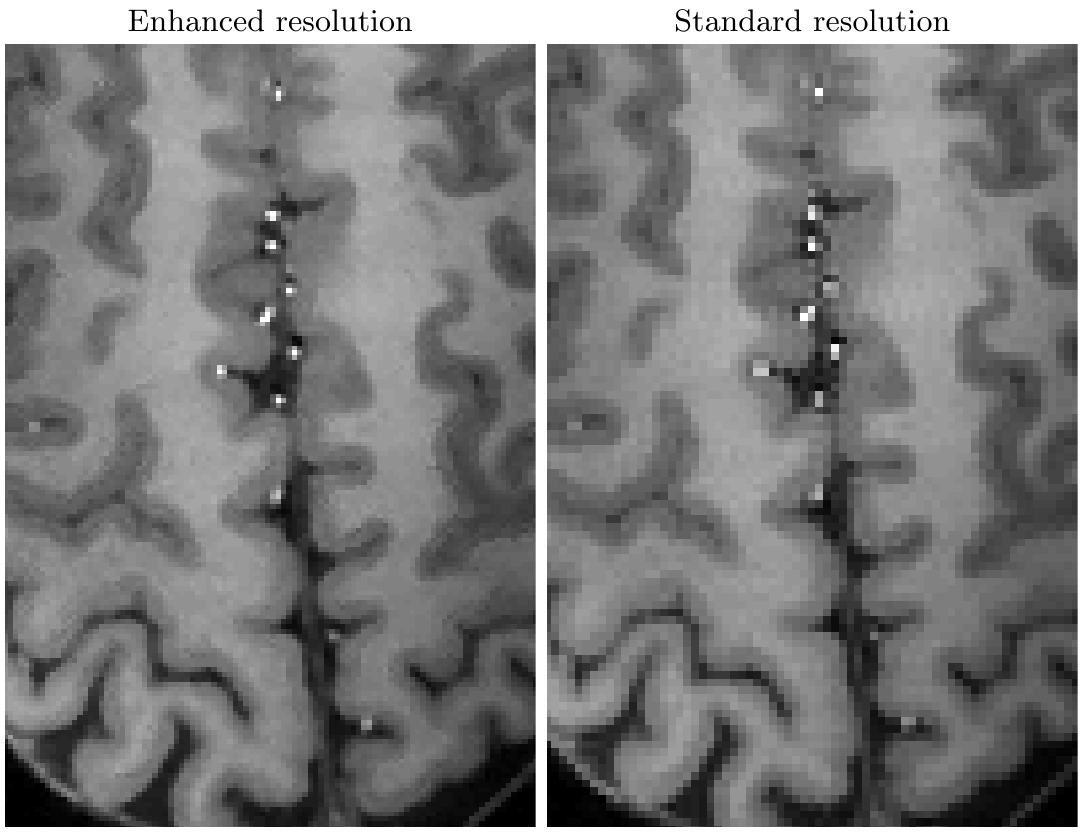}
\caption{\label{fig:high resolution} Axial $(x, y)$ reconstructed slice for a coverage of $100$ percent in the presence of chirp modulation on a high resolution grid of size $N_{\rm c} = 393 \times 255$ (right panel) and after dowsampling on the original grid of size $N = 243 \times 176$.}
\vspace{-7mm}
\end{figure}

Let us also emphasize the potential interest of the s$_2$MRI technique from an enhanced resolution perspective as, in the presence of the chirp modulation, the original image is reconstructed at a high resolution in order to avoid any aliasing problems. One can indeed consider reaching a higher spatial resolution for a fixed acquisition time without probing higher spatial frequencies in practice, which would require stronger gradient coils. In this context, any regularization approach adding a sparsity prior can help to synthesize spatial frequency information higher than that contained in the data. But the chirp modulation implies that high spatial frequency information is actually probed at lower frequencies. For illustration, Fig. \ref{fig:high resolution} shows a reconstructed $(x, y)$ slice for a coverage of $100$ percent on a high resolution grid $N_{\rm c} = 393 \times 255$ as well as the image obtained after downsampling on the grid $N = 243 \times 176$. One can notice that the high resolution image provides sharper details with fewer aliasing artifacts. In particular, the vessels are better resolved.

%
%
% Acknowledgement
%
%
\section*{Acknowledgments}

We thank David Shuman for his helpful suggestions to improve the writing quality of the paper, as well as the reviewers for their very useful comments.

%
%
% BIBLIOGRAPHY
%
%

\end{document}